\documentclass[showpacs,aps,graphicx,twocolumn]{revtex4}
\usepackage{graphicx}
\usepackage{amssymb}
\usepackage{amsmath}
\usepackage{subfigure}
\usepackage{mathrsfs}

\begin{document}

\title{Single phonon source based on a giant acoustic nonlinear effect}
\author{Kang Cai$^1$, Zi-Wen Pan$^3$, Rui-Xia Wang$^1$, Dong Ruan$^1$, Zhang-Qi Yin$^2$\footnote{{Email address:
yinzhangqi@tsinghua.edu.cn}}, Gui-Lu Long$^{1,4,5,}$\footnote{{Email address:
gllong@tsinghua.edu.cn}}}
\affiliation{
$^1$ State Key Laboratory of Low-Dimensional Quantum Physics and Department of Physics, Tsinghua University, Beijing 100084, China\\
$^2$ Center for Quantum Information, Institute for Interdisciplinary Information Sciences, Tsinghua University, Beijing 100084, China\\
$^3$ School of Information and Communication Engineering, Beijing University of Posts and Telecommunications, Beijing, China, 100876\\
$^4$ Tsinghua National Laboratory of Information Science and Technology, Beijing 100084, China\\
$^5$ Collaborative Innovation Center of Quantum Matter, Beijing 100084, China}

\begin{abstract}
We propose a single phonon source based on nitrogen-vacancy (NV) centers, which are located in a diamond
phononic crystal resonator. The strain in the lattice would induce the coupling between the NV centers
and the phonon mode. The strong coupling between the excited state of the NV centers and the phonon is
realized by adding an optical laser driving. This four level NV centers system exhibits the coherent population trapping (CPT), and yields giant resonantly enhanced acoustic nonlinearities, with zero linear susceptibility. Based on this nonlinearity, the single phonon source can be realized.
We numerically calculate $g^{(2)}(0)$ of the single phonon source.  We discuss the effects of the thermal noise and the external driving strength.
\end{abstract}

 \maketitle

\section{Introduction}
In the last decade, one of the most active fields in quantum optics is opto-mechanics, which studies the coupling between the mechanical mode and the optical or microwave field \cite{2014Aspelmeyer_RMP,2012ClerkPRL}. It has wide applications, including gravitational wave detector, squeezing of light, quantum non demolition measurement, etc. The optomechanics also has applications in quantum information processing,  such as an interface between the optical qubits and the superconducting microwaves \cite{2012LinPRL,2014AndrewsNP,2015LinAP}. In its many applications, phonon play a significant role. Thus, the research on phonon has been a hot topic in quantum information processing \cite{2000Rugar,2005RugarNature,2009PainterNature,2010PainterPRL,2012PainterPRL,2015PainterNature,2009RablPRB,2011SeidelinNP,Zhou2014,2014TaoNC,2013LukinPRL,2016HailinWangPRL,2016HailinWangPRX}. Besides above application, there are some unique advantages for phonon in low energy scale. The acoustic wavelength can be as small as $\mu m$ if its frequency is comparable with the microwave photon. The much smaller mode volume on the one hand can realize individual superconducting qubit control \cite{2017SchoelkopfScience}, and on the other hand, can support a large number of modes benefiting for the storage of quantum information. Besides, its potential applications in detection of opaque substances can make up for the disadvantages of optics. Until now, electro-magnetic field induced acoustics transparent has been proposed based on NV center ensemble, which can be used to control phonon velocity in diamond \cite{Hou2016}. Phonon detector is also put forward in recent works \cite{2017HongPhononDetect,2017RXWangPhononDetect}. With more and more attentions focused on phonon, researches on single phonon source becomes indefensible. But the setup proposed now to produce single phonon is either too sophisticated \cite{2008HofheinzNature} or based on measurement \cite{2017HongPhononDetect}.
Thus, a simple and measurement free single phonon source is needed.

The strong nonlinear acoustics interaction is the core for our single phonon source proposal, which is similar to the single photon source \cite{1999YamamotoNature,2005KimbleNature,2008KimbleScience,2008FaraonNP,2011LangPRL,2008LinPRA}. In an optical cavity, the giant nonlinearity can be produced through coupling between optical field and four-level atoms, where one excitation will cause detuning for another excitation \cite{1996SchmidtOL,1997ImamogluPRL,1999ImamogluPRA,1999RebicJOB,2002RebicPRA,2005ImamogluRMP,2006HartmannNP,2007HartmannPRL}. The similar four-level system for phonon can be obtained in the NV centers systems. The NV center in diamond has many advantages, such as the coherence time of the NV centers is very long at room temperature and energy splitting of ground spin states can be adjusted by using magnetic field \cite{2006WrachtrupNV}. The strong coupling between the NV center and the mechanical mode can be realized, either through the magnetic field gradient \cite{2009RablPRB,2017Kang,Ma2016,Ma2017}, or the strain  \cite{2013LukinPRL,2017SchoelkopfScience}. Assisted with optical laser and microwave diving, phonon can be coupled to excited state and ground three spin states of NV centers simultaneously \cite{2016HailinWangPRL,2016HailinWangPRX}, which can be regarded as the effective four-level system. This four-level system that exhibits CPT yields giant resonantly enhanced nonlinearities, while the linear susceptibility is zero. Based on this acoustic nonlinearity, the single phonon state can be produced.

\section{Model}

\begin{figure}[htbp]
\centering
\includegraphics[width=8.5cm,angle=0]{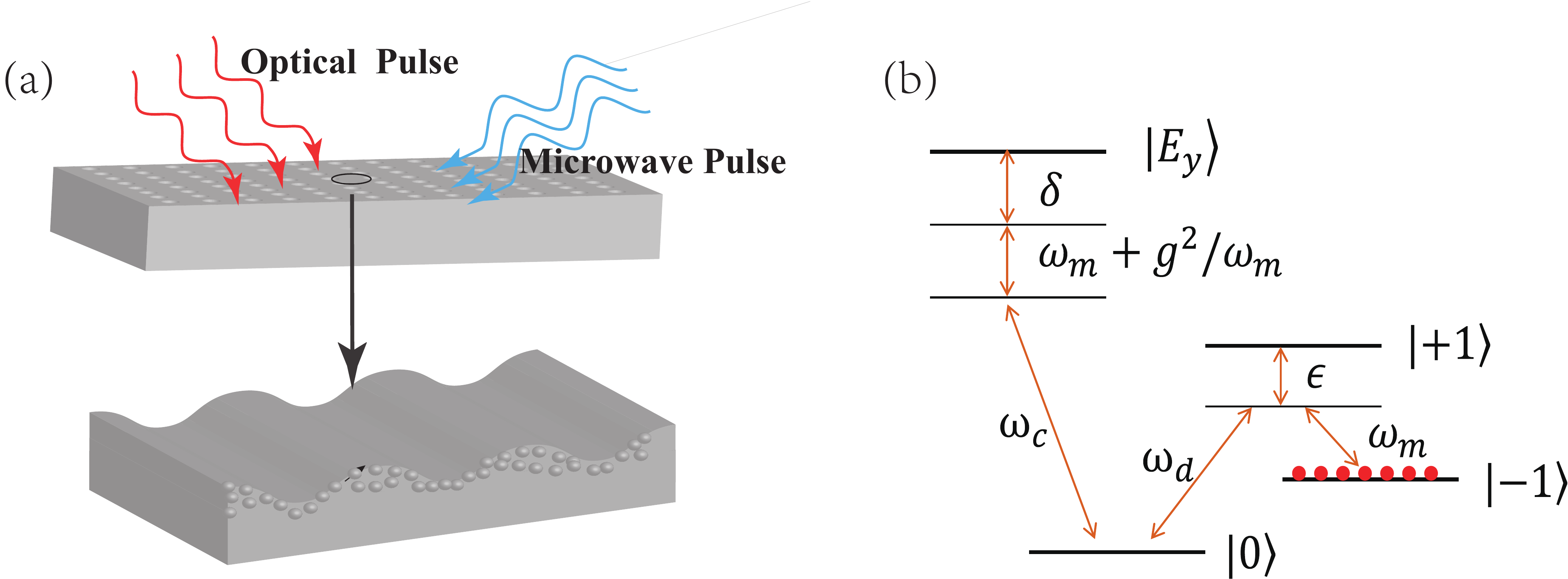}
\caption{(Color online) (a)A schematic diagram of the phononic crystal: The NV center ensembles are located near the surface. Phonon is coupled to NV center ensembles driven by laser field and micro-magnetic field. (b)Energy levels of NV centers. In this structure, we regard $|-1\rangle$,$|0\rangle$,$|+1\rangle$,$|E_{y}\rangle$ as $|1\rangle$,$|2\rangle$,$|3\rangle$ and $|4\rangle$, respectively. $\delta$ and $\epsilon$ are the detuning frequency. $\omega_{m}$, $\omega_{d}$ and $\omega_{c}$ are the frequencies of phonon, optical laser driving and micro-magnetic driving, respectively.}
\label{Model}
\end{figure}

The setup of our model is shown in Fig. \ref{Model}(a). The NV centers ensemble are doped on the surface of phononic crystal made of diamond. The phonon mode in the phononic crystal interacts with the NV centers under the external optical and microwave fields diving. The energy levels and driving in the NV centers are illustrated in Fig. \ref{Model}(b). We focus on the excited state $|E_{y}\rangle$ and ground state $|1\rangle$,$|0\rangle$,$|-1\rangle$ of NV center. An optical field and a phonon mode together are used to dive the transition $|E_{y}\rangle\leftrightarrow |0\rangle$. A microwave field  drives the transition between spin states $|0\rangle$ and $|+1\rangle$ with Rabi frequency $\Omega_d$. The coupling between $|-1\rangle$ and $|+1\rangle$ is magnetic dipole-forbidden. However, it can be induced through phonon mode $b$. For convenience, We re-label spin states $|-1\rangle$,$|0\rangle$,$|+1\rangle$,$|E_{y}\rangle$ as $|1\rangle$,$|2\rangle$,$|3\rangle$,$|4\rangle$, respectively. We define operator $\sigma_{ij}=|i\rangle\langle j|$ and energy difference $\omega _{ij}=\omega _{i}-\omega _{j}$ with $\omega _{i,j}$ frequency of energy level, where $i,j=1,2,3,4$.  The Hamiltonian of the whole system is%
\begin{equation}\label{Original_H}
\begin{aligned}
H_{o} =&\omega _{m}b^{\dagger }b+\omega _{1}\sigma_{11}+\omega _{2}\sigma
_{22}+\left( \epsilon +\omega _{3}\right) \sigma _{33}+\left( \omega
_{4}+\delta \right) \sigma _{44} \\
&+\Omega _{d}\left( e^{-i\omega _{23}t}\sigma _{32}+H.c.\right) +\Omega
_{c}\left( e^{-i\omega _{c}t}\sigma _{42}+H.c.\right)\\
&+g_{24}\left( b^{\dagger }+b\right) \sigma _{44}+g_{13}\left( b\sigma
_{31}+H.c.\right)
\end{aligned}
\end{equation}%
where $b^{\dagger }$ and$\ b$ are the creation and annihilation operators
for the acoustic field with frequency $\omega _{m}$; $g_{13}$ and $g_{24}$ are the
electron-phonon coupling rate; $\omega _{23}$ and $\omega
_{c}$ are the frequency of driving field and optical field with the Rabi frequency of
$\Omega _{d}$ and $\Omega _{c}$, respectively.
Considering the large detuning between optical field and acoustic field, the optical field can be eliminated and its effect can be absorbed into the coupling strength between $|4\rangle$ and $|2\rangle$.
Applying the Schrieffer-Wolff transformation $U=\exp \left[ g_{24}\left( b^{\dagger }-b\right) \sigma_{44}/\omega _{m}\right]$ and under the condition $\omega _{24}-\omega _{c}-g_{24}^{2}/\omega _{m}=\omega
_{m}$, the Hamiltonian in \eqref{Original_H} can be approximated as\cite{2016HailinWangPRL,2016HailinWangPRX}
\begin{equation}
\begin{aligned}
H_{a} \simeq &\ \delta \sigma _{44}+\epsilon \sigma _{33}-\tilde{g}_{24}\left( b\sigma
_{42}+H.c.\right) \\
&+\tilde{g}_{13}\left( b\sigma_{31}+H.c.\right)+\Omega _{d}\left( \sigma _{32}+H.c.\right) ,
\end{aligned}
\end{equation}
where $\tilde{g}_{24}=g_{24}\Omega _{c}/\omega _{m}$. Here, we consider the interaction between surface phonon and NV center ensemble. The coupling strength between phonon $b$ and energy level $\sigma_{31}$ has been enhanced by factor $\sqrt{N}$, namely, $g_{13}\rightarrow \tilde{g}_{13}=\sqrt{N}g_{13}$, with $N$ the number of NV centers.

\section{Nonlinear Effect of Polariton}

The four-level system that exhibits CPT yields giant resonantly enhanced nonlinearities, while the linear susceptibility are identically zero. We divide the Hamiltonian into two parts as $H_{a}=H_{0}+H_{1}$, with
\begin{equation}
\begin{aligned}
H_{0} =&\epsilon \sigma _{33}+\Omega _{d}\left( \sigma _{32}+H.c.\right)
+\tilde{g}_{13}\left( b\sigma _{31}+H.c.\right) , \\
H_{1} =&\delta \sigma _{44}-\tilde{g}_{24}\left( b\sigma _{42}+H.c.\right) .
\end{aligned}
\end{equation}
The first part $H_{0}$ describes the interactions among three spin states of ground state, where three-level $\Lambda $ system is constructed. The second part $H_{1}$ depicts the coupling between excited state and ground state and we analyse this term based on eigenstates of $H_{0}$. We express eigenstates of $H_{0}$ with polariton operators as\cite{2006HartmannNP,2007HartmannPRL}
\begin{equation}\label{Polariton}
\begin{aligned}
P_{0}^{\dagger }&=\left(\tilde{g}_{13}\sigma _{21}-\Omega _{d}b^{\dagger }\right) /B,\\
P_{+}^{\dagger } &=\mu \sigma _{31}+\nu \left( \Omega _{d}\sigma_{21}+\tilde{g}_{13}b^{\dagger }\right) /B,\\
P_{-}^{\dagger } &=-\nu \sigma _{31}+\mu \left( \Omega _{d}\sigma_{21}+\tilde{g}_{13}b^{\dagger }\right) /B,
\end{aligned}
\end{equation}%
where $B=\sqrt{\Omega_{d}^{2}+g^{2}}$ and $\left\vert \mu\right\vert ^{2}+\left\vert \nu \right\vert ^{2}=1$. These operators satisfy the commute formula, $[P_{i},P_{j}^{\dagger }]=\delta_{ij},[P_{i},P_{j}]=[P_{i}^{\dagger },P_{j}^{\dagger }]=0$ with $i,j=0,\pm$, which means that these polaritons are bosons. The polariton operator $P_{0}^{\dagger }$ corresponds to the dark state, which is responsible to the electromagnetically induced transparency.
Using these polaritions, the Hamiltonian $H_{0}$ can be expressed as%
\begin{equation}
H_{0}=\mu _{0}P_{0}^{\dagger }P_{0}+\mu _{+}P_{+}^{\dagger }P_{+}+\mu
_{-}P_{-}^{\dagger }P_{-}
\end{equation}%
where $\mu _{0},\mu _{\pm }=\left( \epsilon \pm A\right) /2$ with $A=\sqrt{\epsilon^{2}+4B^{2}}$. Now, we reconsider the Hamiltonian $H_{1}$ and express the Hamiltonian using the new introduced operators. When condition $\tilde{g}_{24}\Omega _{d}/4\tilde{g}_{13}\ll 2\mu _{+},2\mu _{-},\mu _{+}+\mu _{-}$ is satisfied, the coupling interaction with level-4 can be approximated as%
\begin{equation}\label{CoupleLevel4}
\tilde{g}_{24}\left( \sigma _{41}\sigma _{12}b+b^{\dagger }\sigma
_{21}\sigma _{14}\right) \simeq -\frac{\tilde{g}_{24}\tilde{g}_{13}}{2\Omega _{d}}%
\left[ \sigma _{41}\left( P_{0}\right) ^{2}+\left( P_{0}^{\dagger }\right)
^{2}\sigma _{14}\right].
\end{equation}%
Of all parameters, $\Omega _{d}$ is most easy to adjust. In order to obtain \eqref{CoupleLevel4}, it seems that we can adjust $\Omega _{d}$ infinite small, but this is not the case. The higher-order nonlinearity of $P_{0}$ requires ${\tilde{g}_{24}\tilde{g}_{13}}/{2\Omega _{d}}\ll \delta$, and only then, energy shift for the level $|4\rangle$ can be obtained using the $2nd$ perturbation theory as\cite{1999RebicJOB,2002RebicPRA,2005ImamogluRMP,2006HartmannNP,2007HartmannPRL}
\begin{equation}\label{Nonliear_H}
H_{E}=-g\left(
P_{0}^{\dagger }\right) ^{2}\left( P_{0}\right) ^{2},
\end{equation}%
which is the effective giant nonlinear effect of dark polariton operator $P_{0}^{\dagger }$ with $g={\tilde{g}_{24}^{2}\tilde{g}_{13}^{2}}/{4\delta \Omega _{d}^{2}}$, the effective coupling strength. The exist of one polariton in the system will blockade absorbing of the second polariton. In this process, the dissipation of polariton $P_{0}^{\dagger }$ should also be considered. The dissipation mainly includes two parts, $\gamma=\gamma_{a}+\gamma_{p}$. $\gamma_{a}$ comes from the spontaneous decay of the level $|4\rangle$, which can estimated as $\gamma_{a} =(\gamma_{4}/2\delta)g$. $\gamma_{p}$ is caused by the dissipation of phonon in phononic crystal. In this part, we can see that $\delta$ should be large enough to suppress the dissipation $\gamma_{a}$ of polariton $P_{0}^{\dagger }$ and however, the nonlinear strength $g$ requires that the value of $\delta$ should be small. Therefore, choosing balanced parameters is important in our scheme.

\section{Single Polariton and Phonon Preparation}

In order to produce obvious large nonlinear effect, small value of parameter $\Omega_{d}$ is required and thus, only single polartion $P_{0}^{\dagger}$ is prepared at first. Then, adjusting parameter $\Omega_{d}$ adiabatically \cite{2017ShaoPRA,1932Zener} until $\Omega _{d}>>\tilde{g}_{13}$, the polariton will evolves into the form of phonon according to the form of dark state polariton $P_{0}^{\dagger}$ as Eq. \eqref{Polariton}. Now, we first focus on preparation of single dark state polartion $P_{0}^{\dagger}$. The microwave is introduced to driving the transition between states $|1\rangle$ and $|2\rangle$ of NV centers with strength $\Omega$. Since $\sigma_{21}=\Omega_{d}(\nu^{2}P_{+}^{\dagger}+\mu^{2}P_{-}^{\dagger})/B+\tilde{g}_{13}P_{0}^{\dagger}/B$, the driving $\Omega$ is actually applied to driving polaritons. We choose parameters to satisfy conditions $\Omega_{d}<<\tilde{g}_{13}$ and $\Omega<{\tilde{g}_{24}^{2}\tilde{g}_{13}^{2}}/{4\delta \Omega _{d}^{2}}<< \mu _{\pm}$, which guarantees  the driving effect on polaritons $P_{\pm}$ can be neglected and the driving is mainly to excited dark state polariton $P_{0}^{\dagger}$. The Hamiltonian is
\begin{equation}
H=-g\left(
P_{0}^{\dagger }\right) ^{2}\left( P_{0}\right) ^{2}+\tilde{\Omega} (P_{0}^{\dagger }+P_{0}),
\end{equation}%
where $\tilde{\Omega}=\Omega\tilde{g}_{13}/B$, is driving strength applied on dark state polariton $P_{0}^{\dagger}$. The statistical properties of the single phonon sources can be measured through the second-order correlation function,
\begin{equation}
g^{(2)}(0)={\langle P_{0}^{\dagger}(t)P_{0}^{\dagger}(t)P_{0}(t)P_{0}(t)\rangle}/{\langle P_{0}^{\dagger}(t)P_{0}(t)\rangle^2}.
\end{equation}
$g^{(2)}(0)>1$ means that radical sources tend to emit polaritons in bunches with super-Poisson distributed statistic while $g^{(2)}(0)<1$ indicates that emitting polaritons are one by one well separated in time from each other with antibunching,a unique quantum characteristic of the field.

To simulate this process, the influence of environment must be considered. The state of our system is usually in the mixed state, expressed by density matrix $\rho(t)$ and its dynamical process can be depicted as
\begin{equation}\label{masterequation}%
\begin{aligned}
\dot{\rho}\left( t\right)=&-i\left[ H,\rho \left( t\right) \right]\\
 &+\frac{\gamma}{2}(1+n_{th}) L\left[ P_{0}\right]
\rho \left( t\right)+\frac{\gamma}{2} n_{th} L\left[ P_{0}^{\dagger}\right]
\rho \left( t\right),
\end{aligned}
\end{equation}
with $L\left[ o\right] \rho=2o\rho o^{\dagger }-o^{\dagger }o\rho -\rho o^{\dagger }o$, $\gamma$ the dissipation of the system, $n_{th}=\langle P_{0}^{\dagger}P_{0}\rangle$ the mean thermal polariton number. The initial polariton state is in the thermal state $\rho(0)=\sum_{n=0}^{\infty}p_{n}|n\rangle\langle n|$, where $p_{n}=n_{th}^{n}/(1+n_{th})^{n+1}$ is the probability of state $|n\rangle$. We initialize all the NV centers in level $|1\rangle$, the mean thermal number of polariton $P_{0}^{\dagger}$ only depends on the mean thermal phonon, namely, $\langle P_{0}^{\dagger}P_{0}\rangle=\Omega_{d}^{2}\langle b^{\dagger}b\rangle/B^{2}$.

\begin{figure}[htbp]
\centering
\includegraphics[width=9cm,angle=0]{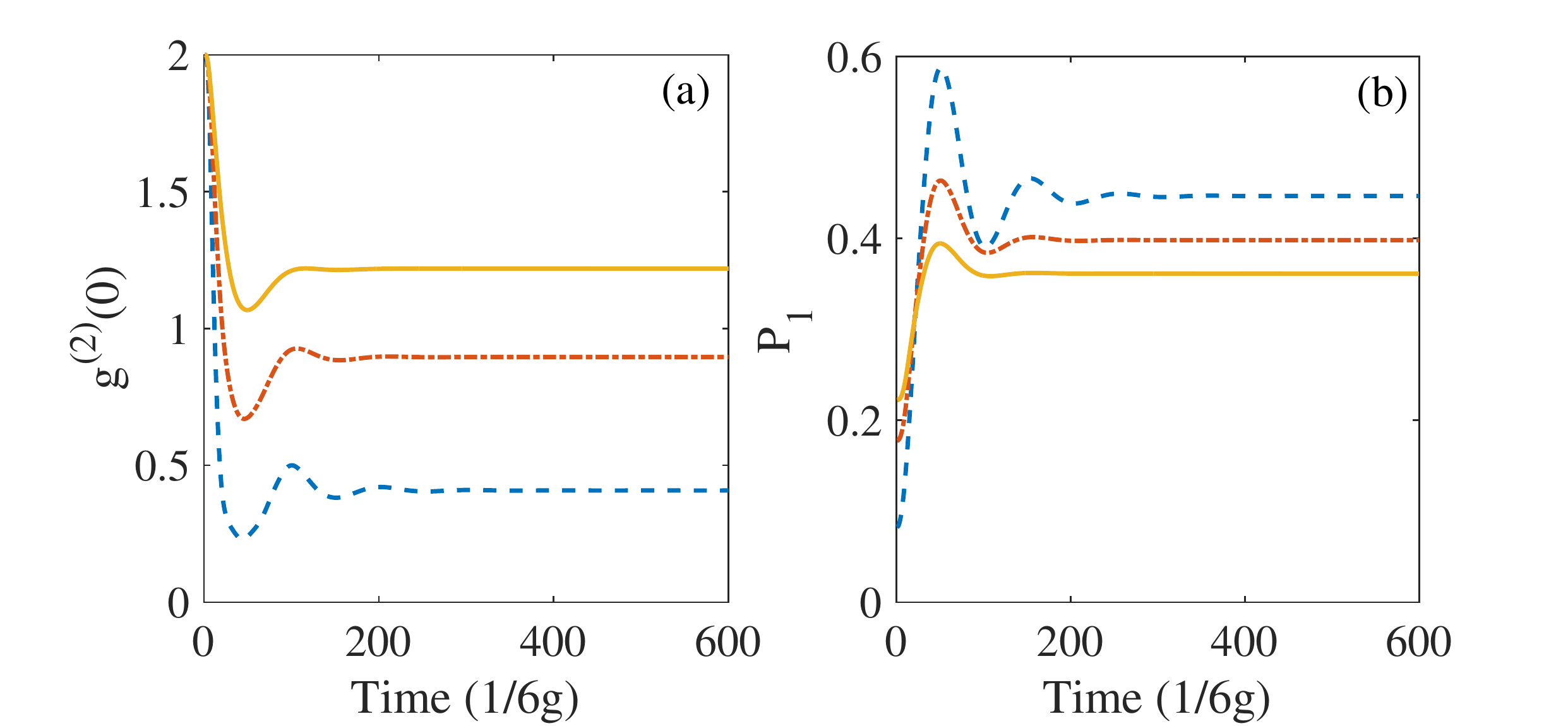}
\caption{(Color online) (a)Evolution of the second correlation function $g^{(2)}(0)$; (b)Evolution of the population of Fock state $|1\rangle$). The coupling strength $g/2\pi=25kHz$, the dissipation of system $\gamma=g/8$, the driving strength is $\tilde{\Omega}=g/5$. The mean thermal polariton number of blue dash line, red dot dash line and yellow line are $\langle P_{0}^{\dagger}P_{0}\rangle=0.1,0.3,0.5$, respectively.}
\label{Result1_thermalnoise}
\end{figure}

Now, we will present numerical simulation of dynamical process as \eqref{masterequation}. We choose the parameters as $N=40000$, $g_{13}/2\pi=1kHz$ \cite{2013LukinPRL}, $\Omega _{d}/2\pi=20kHz$, $\tilde{g}_{24}/2\pi=200kHz$, $\epsilon/2\pi=200kHz$, $\delta/2\pi=40MHz$, $\omega_{m}/2\pi=800M$. The strength of nonlinear interaction can be calculated as $g/2\pi=25kHz$ immediately. At temperature about $0.5K$, the mean number of polariton is $\langle P_{0}^{\dagger}P_{0}\rangle=0.1$. As for the dissipation $\gamma$ of polariton,  $\gamma_{4}/2\pi=10MHz$ yields $\gamma_{a}=g/8$ and $\gamma_{p}/2\pi=800Hz$ with the quality factor of phononic crystal $Q=10^{6}$ \cite{2017SchoelkopfScience}. Since $\gamma_{p}=\omega_{m}/Q<<\gamma_{a}$, the dissipation $\gamma$ of polariton can be approximated as $\gamma=\gamma_{a}$. Hilbert space is chosen as $\{|n\rangle\}_{n=0}^{n_{max}}$, where $|n\rangle$ is the Fock state of polaritons and $n_{max}=20$ is the upper cutoff in our calculation.

\begin{figure}[htbp]
\centering
\includegraphics[width=9cm,angle=0]{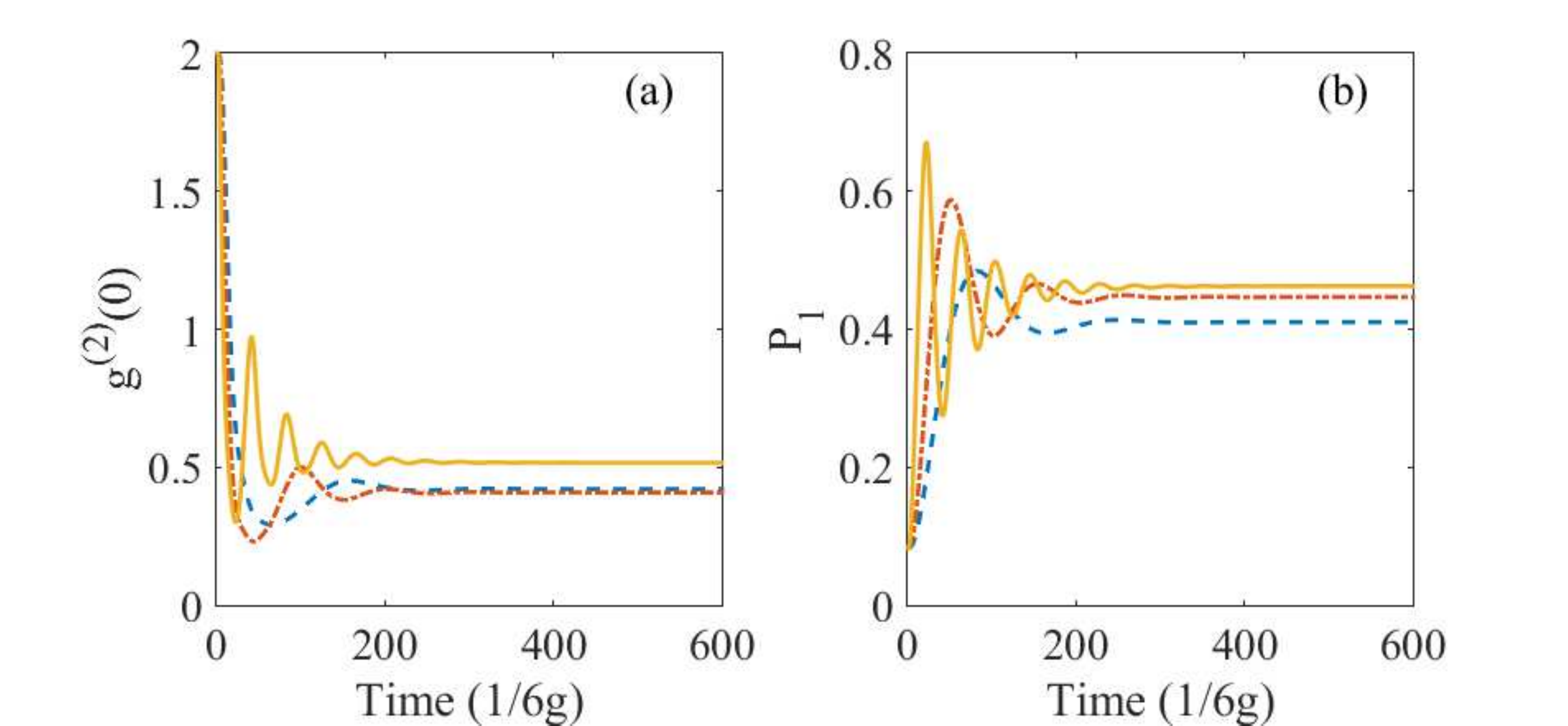}
\caption{(Color online) (a)Evolution of the second correlation function $g^{(2)}(0)$; (b)Evolution of the population of Fock state $|1\rangle$). The coupling strength $g/2\pi=25kHz$, the dissipation of system $\gamma=g/8$, the mean thermal polariton number is $\langle P_{0}^{\dagger}P_{0}\rangle=0.1$. The driving strength of blue dash line, red dot dash line and yellow line are $\tilde{\Omega}=g/8,g/5,g/2$, respectively.}
\label{Result2_driving}
\end{figure}

The second-order correlation function $g^{(2)}(0)$ and the population of Fock state $|1\rangle$ $P_{1}$ are calculated, and shown in Fig. \ref{Result1_thermalnoise} and Fig. \ref{Result2_driving}. The initial state is in thermal state, corresponding to $g^{(2)}(0)=2$. With time going on, $g^{(2)}(0)$ decreases until $g^{(2)}(0)<1$, which means that the statistic of our photon source changes from super-Poisson distribution to sub-Poisson distribution and Fock state $|1\rangle$ becomes dominated. When $g^{(2)}(0)$ evolves to the minimum point, $P_{1}$ reaches its peak and the reason for this is obvious that the minimal $g^{(2)}(0)$ represents the maximal probability to produce the single polariton. The thermal state will demolish the classical properties of our system, which is shown in Fig. \ref{Result1_thermalnoise} where the minimal value of $g^{(2)}(0)$ increases with increasing of mean thermal number of polariton. Also, from Fig. \ref{Result2_driving}, we can see that the time for $g^{(2)}(0)$ to reach the minimal value will decrease when the driving strength increases. Thus, the increasing of the driving strength can save the time to obtain the single polarition which is useful to resist decoherence.

When the minimum point of $g^{(2)}(0)$ is reached, turn off the driving $\tilde{\Omega}$ immediately. The Fock state $|1\rangle$ of polariton $P_{0}^{\dagger}$ dominates at this time. Then, we adjust the driving strength of microwave $\Omega _{d}\exp(vt)$ with velocity $v$. The velocity should satisfy the adiabatic condition $v\ll\mu _{+},\mu _{-}$, which makes sure polaritions $P_{\pm}$ cannot be excited
in this process \cite{2017ShaoPRA,1932Zener}. The dissipation mainly comes from the dissipation of phonon $\gamma_{p}$ and the evolution time is limited by $1/\gamma_{p}$. The proportion of phonon increases with the increasing of $\Omega _{d}(t)=\Omega _{d}\exp(vt)$. When $\Omega _{d}(t)>>\tilde{g}_{13}$, this polariton transforms into phonon, which is the essence of single phonon source. In our proposal, we choose $v=\tilde{g}_{13}/5$ and at time $t=25/\tilde{g}_{13}$, $\Omega _{d}(t)=\Omega _{d}\exp(5)>>\tilde{g}_{13}$ , meanwhile, $t<<1/\gamma_{p}$. Therefore, a single phonon source can be realized based on our scheme.

\section{Conclusion}

In conclusion, we have proposed a scheme to produce a single phonon based on the nonlinear effect in interactive process between the phonon and the NV centers. We have shown that the nonlinear coupling strength can be stronger than the phonon decay rate. We have also calculated the second correlation function $g^{(2)}(0)$ numerically, and found that $g^{(2)}(0)<1$ for practical parameters. Finally, the effect of thermal noise and external driving strength on $g^{(2)}(0)$ has been simulated and discussed. Recently, the researches on phonon has witnessed significant progresses and we hope that our study stimulates further experimental researches on the applications of phonon in quantum information processing.

\section*{Acknowledgments}
We acknowledge the financial support from the National Natural Science Foundation of China under Grant No. 20141300566, 61435007 and 61771278. Rui-Xia Wang acknowledges the support from China Postdoctoral Science
Foundation under Grant No. 2016M600999.

\end{document}